# Electromagnetic-Polarization Selective Composites with Quasi-1D van der Waals Metallic Fillers

Zahra Barani[1], Fariborz Kargar[1,2*], Yassamin Ghafouri[3], Saba Baraghani[1,2], Sriharsha Sudhindra[1], Amirmahdi Mohammadzadeh[1], Tina T. Salguero[3], and Alexander A. Balandin[1,4*]

[1]Nano-Device Laboratory (NDL) and Phonon Optimized Engineered Materials (POEM) Center, Department of Electrical and Computer Engineering, University of California, Riverside, California 92521 USA

[2]Department of Chemical and Environmental Engineering, University of California, Riverside, California 92521 USA

[3]Department of Chemistry, University of Georgia, Athens, Georgia 30602 USA

[4]Material Science and Engineering Program, University of California, Riverside, California 92521 USA

## Abstract

We report on the preparation of flexible polymer composite films with aligned metallic fillers comprised of atomic chain *bundles* of the quasi-one-dimensional (1D) van der Waals material tantalum triselenide, $TaSe_3$. The material functionality, embedded at the nanoscale level, is achieved by mimicking the design of an electromagnetic aperture grid antenna. The processed composites employ chemically exfoliated $TaSe_3$ nanowires as the grid building blocks incorporated within the thin film. Filler alignment is achieved using the "blade coating" method. Measurements conducted in the X-band frequency range demonstrate that the electromagnetic transmission through such films can be varied significantly by changing the relative orientations of the quasi-1D fillers and the polarization of the electromagnetic wave. We argue that such polarization-sensitive polymer films with quasi-1D fillers are applicable to advanced electromagnetic interference shielding in future communication systems.

**Keywords:** quasi-1D van der Waals materials; polymer composites; electromagnetic properties

---

* Corresponding authors: fkargar@ece.ucr.edu (F.K.); balandin@ece.ucr.edu (A.A.B.); web-site: http://balandingroup.ucr.edu/

Commonly, one selects functional materials with known characteristics to build a device or a system. In more elaborate approaches, one can engineer and synthesize materials with the required properties for specific applications. The inspiration for material selection, composition, and assembly can come from diverse sources. In one well-known approach – biomimetics – the models and elements of nature are applied to the design of synthetic systems.[1,2] In an analogous approach, well-developed design solutions for macroscopic objects are translated into micro-, nano-, or atomic-scale structures. We followed this innovative path to create a polymer composite with polarization-sensitive electromagnetic interference (EMI) shielding characteristics by emulating the macroscopic structure and, to some degree, the functionality of an electromagnetic (EM) grid aperture antenna at the nanoscale level. A polarization-selective grid antenna is a set of parallel metal grid lines that allow transmission or reflection of radio-frequency (RF) radiation depending on the polarization of the radiation.[3] This design allows a single structure to act as a mirror for RF radiation or become transparent to such radiation. When the polarization of the electric field is parallel to the grid lines, the electric field induces a current in the grid lines, which reflects the EM wave. In the alternate case, with the polarization of the electric field perpendicular to the grid lines, no current is induced, and the EM radiation passes through the grid. Polarization selection grids are often manufactured with metal wire tracks, usually copper, on a dielectric substrate. The spacings between grid lines must be small relative to the wavelength of the linear polarized EM waves. Here we use a similar antenna design, albeit at the nanometer scale, to create a "grid-antenna film."

In this work, we mimic the grid antenna design in nano-composites by employing chemically-exfoliated *bundles* of a quasi-one-dimensional (1D) van der Waals material, tantalum triselenide (TaSe$_3$). We recently demonstrated the potential of TaSe$_3$ for extremely high current density[4–6] and effective EMI shielding, even with random filler distribution and low filler loading fractions.[7] Such quasi-1D van der Waals materials are less well-investigated compared with two-dimensional (2D) layered van der Waals materials, such as graphene and transition metal dichalcogenides (TMDs).[8–10] The quasi-1D van der Waals materials include the transition metal trichalcogenides (TMTs) with formula MX$_3$ (M = transition metal, X = S, Se, Te), such as TiS$_3$, NbS$_3$, TaSe$_3$, and ZrTe$_3$, as well as other materials containing 1D structural motifs.[11–13] As opposed to TMDs,

TMTs exfoliate into nanowire- or nanoribbon-type structures,[4,12–15] which stem from their unique chain-based crystal structures, illustrated for TaSe$_3$ in Figure 1a. In principle, these low dimensional materials can be exfoliated into individual atomic chains or few-chain atomic threads. Theory suggests that there are many quasi-1D van der Waals materials that retain their metallic or semiconductor properties when exfoliated to atomic chains.[16–18] The exfoliated bundles of TaSe$_3$ atomic threads with cross-sections in the range of 10 nm × 10 nm to 100 nm × 100 nm revealed exceptionally high current densities of up to ~ 30 MA/cm$^2$, an order of magnitude higher than that of copper.[19,20] Additionally, the liquid phase exfoliated (LPE) TaSe$_3$ bundles can be millimeters in length, providing substantial aspect ratios. The current-carrying capability of the metallic TaSe$_3$, in addition to their high-aspect-ratios, allow us to use them as "metallic grids" even when scaled down to 100-nm features or below.

[Figure 1: Crystal structure of quasi-1D TaSe$_3$ used in this study.]

For this study, we used TaSe$_3$ crystals prepared by chemical vapor transport (CVT). In contrast to typical CVT reactions, where the goal is the growth of a relatively small number of larger crystals,[4,5,15] here we varied reaction conditions to yield 0.7-1.5 g batches of TaSe$_3$ crystals for composite preparation. Transport was achieved using iodine as the transport agent and/or by using a 625 °C – 600 °C temperature gradient. As can be seen in Figure 1b, the scale of these reactions led to the growth of TaSe$_3$ crystals almost entirely filling the ampule volume. The resulting mat of crystals could be removed easily, providing 18 % - 38 % yields of mm- to cm-long needlelike or fibrous crystals (Figure 1b), and also leaving behind a quantity of microcrystalline solid that was not used for subsequent exfoliation. The long crystals have the smooth faces and straight edges that characterize high-quality TaSe$_3$ samples (Figure 1c-d and Figure S1). Powder x-ray diffraction (Figure S2) and energy dispersive spectroscopy (see Table S1 for compositional analysis) provided analytical results consistent with the standard structure of TaSe$_3$, albeit with some variation in Se content.

These CVT grown TaSe$_3$ crystals were subjected to the LPE following the process reported by us elsewhere.[7,21] Figure 2 summarizes our approach of mimicking the grid antenna design using material synthesis and presents optical images of the partially aligned fillers in the polymer composites and the resulting films. We used the "blade coating" method to prepare flexible thin films with a thickness of 100 µm ± 10 µm with a special type of UV-cured polymer and exfoliated TaSe$_3$ as fillers. In this method, a small amount of polymer-filler solution is drop cast on a rigid substrate with a smooth surface.[22,23] A blade with an adjustable distance from the top surface of the substrate is gradually run over the mixture and spread the compound over the substrate (Figure 2b). Using this technique, the quasi-1D fillers are aligned, to some extent, in the direction of the coating owing to the applied viscoelastic shear stress as a result of blade movement[22,23] (Figure 2c). The samples are referred to as A, B, C, and D throughout this manuscript with filler concentrations of 2.2 vol%, 1.03 vol%, 1.87 vol%, and 1.61 vol%, respectively. The properties of the samples are summarized in Table S2 of Supporting Information.

[Figure 2: System-level concept and material-level implementation.]

We conducted EM testing of the prepared films in the X-band frequency range ($f =$ 8.2 GHz – 12.4 GHz), which is pertinent to the current and future communication technologies. To determine the polarization selectivity, we followed the measurement protocols used in EMI shielding testing.[24–28] We measured the scattering parameters, $S_{ij}$, with the two-port programmable network analyzer (PNA; Keysight N5221A). The scattering parameters are related to the coefficients of reflection, $R = |S_{11}|^2$, and transmission, $T = |S_{21}|^2$. The measurements were carried out in a WR-90 commercial grade straight waveguide with two adapters at both ends with SMA coaxial ports. The samples were made a bit larger than the rectangular cross-section ($a =$ 22.8 mm, $b =$ 10.1 mm) of the central hollow part of the waveguide to prevent the leakage of the EM waves from the sender to the receiver antenna. The cut-off frequency for different fundamental transverse electric (TE) modes in rectangular shaped waveguides is $(f_c)_{mn} = \frac{1}{2\pi\sqrt{\mu\varepsilon}}\sqrt{(m\pi/a)^2 + (n\pi/b)^2}$ [Hz] where $m$ and $n$ are positive integer numbers.[29] Therefore, the dominant EM mode in WR-90 waveguide is TE$_{10}$ mode with electrical field ($E$) oscillating in the vertical direction perpendicular to the larger side of the inlet aperture (see Figure 3). The frequencies of other modes exceed the X-band frequency range and are not of interest in this study.

[Figure 3: Schematic of EM measurements.]

To investigate the effect of the filler alignment on the EM characteristics of the composites, measurements were carried out at different sample orientation angles ($\alpha$) by rotating the sample about the guide axis. Note that $\alpha$ is the angle between the aligned filler chains in the composite with respect to the larger side of the guide's aperture. Therefore, at $\alpha = 0°$, the fillers are parallel to the larger side and $E$ is perpendicular to them. The front-view schematic of the WR-90 waveguide, the electric field configuration of $TE_{10}$ mode, and its mutual orientation with respect to the quasi-1D fillers of the composites are shown in Figure 3. Figure 4 presents the reflection ($SE_R$), absorption ($SE_A$), and total ($SE_T$) shielding effectiveness of samples A and C with 2.2 vol% and 1.87 vol% filler concentration as a function of EM frequency when the polarization of the incident EM wave is either parallel with (∥) or transverse to (⊥) the quasi-1D fillers. Note that the shielding effectiveness of the films is significantly enhanced when $E$ is parallel with the filler alignment comparing to the case it is perpendicular to the filler chains.

[Figure 4: EM shielding with respect to the orientation of the atomic chains.]

We measured the angular dependence of $SE_T$ of all four samples at the constant frequency of $f = 8.2$ GHz to elucidate the effect of the filler alignment on EM shielding properties of the composites. The results of these measurements are presented in Figure 5. Note that at $\alpha = 0°$, E is perpendicular to the filler atomic chains. One can notice the sinusoidal characteristic of the $SE_T$ (solid lines) with a period of 180 degrees as a function of $\alpha$. When the fillers are aligned in one direction, the composite film becomes anisotropic, given that the embedding matrix is isotropic. The results are consistent for all examined samples with different loading of the aligned quasi-1D fillers. The shielding effectiveness of the samples prepared by the "compressional molding," with randomly oriented fillers, does not exhibit any angular dependency characteristic (see Supplementary Figure S8 and S9). The latter provides additional evidence of the filler alignment when the "blade coating" method is applied during the sample synthesis.

[Figure 5: Angular dependency of the shielding effectiveness.]

The periodic EM shielding characteristics observed in our composites originate from two different effects: (i) the prolate ellipsoidic needle-like geometry of the fillers, assuming semi axes of $a_x > a_y = a_z$; and (ii) the anisotropic complex permittivity properties of the quasi-1D TaSe$_3$ fillers.[30] Because the filler inclusion is low in all the samples, the Maxwell-Garnett (M-G) effective medium theory can be used to explain the observed characteristics.[30,31] For composites with aligned dielectric fillers, the M-G effective complex permittivity of the composite along the $x$ direction, $\varepsilon_{c,x}$, can be described as:

$$\varepsilon_{c,x} = \varepsilon_p + \phi_f \varepsilon_p \frac{\varepsilon_f - \varepsilon_p}{\varepsilon_f + (1 - \phi_f) N_x (\varepsilon_f - \varepsilon_p)} \qquad (1)$$

In this equation, $\varepsilon_p$ and $\varepsilon_f$ are the permittivity of the polymer and filler and $\phi_f$ is the filler volume fraction. For ellipsoidal fillers, the depolarization factor, $N_x = (\frac{1-e^2}{2e^3})(\ln\frac{1+e}{1-e} - 2e)$ in which the eccentricity is $e = \sqrt{1 - a_y^2/a_x^2}$.[30] Considering the large aspect ratio of the exfoliated fillers ($a_x \gg a_y$), it can be easily inferred that the effective permittivity of the composites would be largely different along different directions, i.e. parallel with and and perpendicular to the aligned atomic chains. Note that to obtain the effective permittivity along other directions, $y$ and $z$, the depolarization factor should be replaced by $N_y$ and $N_z$ and calculated accordingly.

The special geometrical shape of the aligned fillers is not the only parameter causing anisotropic behavior of compoistes with quasi-1D fillers. Owing to the monoclinic crystal structure of TaSe$_3$, the EM properties of the fillers are highly directional. The polarized reflectance data of TaSe$_3$ exhibits metallic characteristic in the infrared region.[32] To the best of our knowledge, there is no data on the dielectric properties of TaSe$_3$ in the microwave region. However, one can describe the complex dielectric parameterof TaSe$_3$ as a function of EM frequency, $\omega$, by the Lorentz-Drude model, $\varepsilon(\omega) = \varepsilon_\infty - \frac{\omega_p^2}{\omega^2 - i\omega\Gamma_0} + \sum_{n=1}^{m} \frac{\omega_{p,n}^2}{\omega_{0,n}^2 - \omega^2 - i\omega\Gamma_n}$.[32] In this model, $\varepsilon_\infty$ is the permittivity of the material when the frequency goes to infinity, $m$ is the number of the oscillators with the frequency of $\omega_{0,n}$ and the lifetime of $1/\Gamma_n$, respectively. The plasma frequency, $\omega_p = \sqrt{Nq^2/m^*\varepsilon_0}$, depends on the electron density, $N$, electron absolute charge, $q$, and the effective mass of electrons, $m^*$. The second and third terms are associated with the interaction of EM waves with the intra-band, or free-electrons, and inter-band, or bound-electrons, respectively. The $\hbar\omega_p$ in TaSe$_3$ along the crystallographic "a" (perpendicular to the chains) and "b" (along the chains) axes are 0.42 eV and 0.68 eV.[32] Therefore, one would expect an anisotropic frequency-dependent reflectance (R) and conductance ($\sigma$) along with and perpendicular to the atomic chains in the microwave region, with both parameters being larger in the direction along the atomic chains. Such strong, anisotropic reflectance properties have been reported for TaSe$_3$ in the EM energy range between 0.05 eV to 5 eV, previously. [33]

Figure 6 (a-b) exhibits the angular dependent reflection, absorption, and transmission shielding effectiveness and coefficients of sample D, respectively. The four-fold symmetry of all plots shown in both panels confirms the alignment of quasi-1D fillers. More importantly, as it is seen in Figure 6 (b), reflection is the dominant mechanism of shielding of the EM waves in the microwave region. The reflection coefficient increases more than two times comparing the two extreme cases of $\alpha = 0°$ and $\alpha = 90°$ whereas the absorption almost does not vary. The four-fold symmetric transmission curve in Figure 6 (b) demonstrates the applicability of prepared flexible films as microwave attenuators or grid polarizers. For reference, the shielding effectiveness in the samples prepared by the "compressional molding" does not reveal any angular dependence (see Supplementary Figure S7-a and S8). The observed EM property is similar to the linear dichroism, which has been reported in the visible light region for bulk and exfoliated MPX$_3$ crystals with strong optical anisotropy.[34] Bulk TiS$_3$ exhibit a linear dichroism with transmittance ratio of $\zeta = T_\perp/T_\parallel = 30$ at the wavelength of 633 nm.[35]

[Figure 6: Contribution of various mechanisms to EM waves interactions with the composites.]

In conclusion, we have described the preparation and properties of flexible polymer composite films with aligned metallic fillers made of bundles of quasi-one-dimensional (1D) van der Waals metal, characterized by high current density. The material functionality, embedded at the nanoscale level, was achieved by mimicking the design of an electromagnetic aperture grid antenna. The synthesized composites use the quasi-1D van der Waals nanowires as the grid building block incorporated within the thin-film structure. The measurements conducted in the X-band frequency range demonstrated that the electromagnetic transmission through such films could be varied by changing the mutual orientation of the quasi-1D fillers and polarization of the electromagnetic wave. The films with low loading of the quasi-1D fillers (< 2 vol. %) and only partial alignment of the fillers can already produce ~5 dB variation in the transmitted signal. We argue that such polarization-sensitive polymer films with quasi-1D fillers can be used for advanced electromagnetic interference shielding in future communication systems.

**Acknowledgments**

The work at UC Riverside was supported, in part, by the National Science Foundation (NSF) program Designing Materials to Revolutionize and Engineer our Future (DMREF) *via* a project DMR-1921958 entitled Collaborative Research: Data Driven Discovery of Synthesis Pathways and Distinguishing Electronic Phenomena of 1D van der Waals Bonded Solids.

**Contributions**

A.A.B. and F.K. conceived the idea of the EMI shielding films with oriented quasi-1D van der Waals fillers, planned the study and led the manuscript preparation. Z.B. developed the testing

**Captions**

**Figure 1: Crystal structure of quasi-1D TaSe$_3$ used in this study.** (a) Crystal structure of TaSe$_3$ (blue = Ta, red = Se) with two views illustrating interchain interactions and emphasizing this material's 1D nature originating from chains extending along the b-axis. (b) Schematic of the CVT process employed here to prepare TaSe$_3$ crystals (top) and a photograph of an as-synthesized mass of crystals removed from its growth ampule (below). (c) SEM image of TaSe$_3$ crystals highlighting their high aspect ratio. (d) Secondary electron (SE) image of a TaSe$_3$ nanowire produced by solvent exfoliation.

**Figure 2: System-level concept and material-level implementation.** (a) View of the aperture grid antenna illustrating the required function – polarization selectivity. (b) Schematic of the "blade coating" filler alignment process in the polymer films, in which the bundles of quasi-1D atomic threads function as metal wires in a grid antenna. (c) Optical microscopy image of the UV-cured polymer film with 1.8 vol% of TaSe$_3$ quasi-1D fillers. Note the aligned high-aspect-ratio TaSe$_3$ fillers along the coating direction. (d) Optical image of the resulting flexible polymer films with incorporated quasi-1D fillers, which mimic the action of a grid antenna.

**Figure 3: Experimental procedures.** (a) Front-view schematic of the standard WR-90 waveguide and polarization of the allowed fundamental TE$_{10}$ mode propagating in this type of waveguide at a given frequency range. At $\alpha = 0°$, the bundles of the quasi-1D atomic chains are parallel to the large side of the aperture and perpendicular to the electric field of TE$_{10}$ mode.

**Figure 4: Electromagnetic data.** Reflection (SE$_R$), absorption (SE$_A$), and total (SE$_T$) EMI shielding effectiveness of (a-b) sample A and (c-d) sample C for two cases of the EM wave polarization transverse to ($\perp$) and parallel with ($\parallel$) the quasi-1D fillers. Note that the EMI shielding is significantly higher when the polarization is parallel to the filler alignment.

**Figure 5: Electromagnetic-polarization selective composites.** Total shielding effectiveness of all samples as a function of the composite orientation angle $\alpha$, measured at 8.2 GHz frequency. At $\alpha = 0°$, the polarization of the EM wave is perpendicular to the filler alignment. The results are fitted with sine functions. Note the periodicity of the $SE_T$ with a period of 180 degrees.

**Figure 6: Contribution of different mechanisms to interaction with EM waves.** Angular dependency of (a) the reflection, absorption, and total shielding effectiveness, and (b) reflection absorption, and transmission coefficients of sample D with 1.61 vol% aligned quasi-1D $TaSe_3$ fillers. Note the extremes at 0 and 90 degrees and the symmetry of the curves in both panels confirming the alignment of fillers inside the polymer matrix. As shown in (b) the reflection is highly correlated with sample orientation whereas absorption varies weakly.

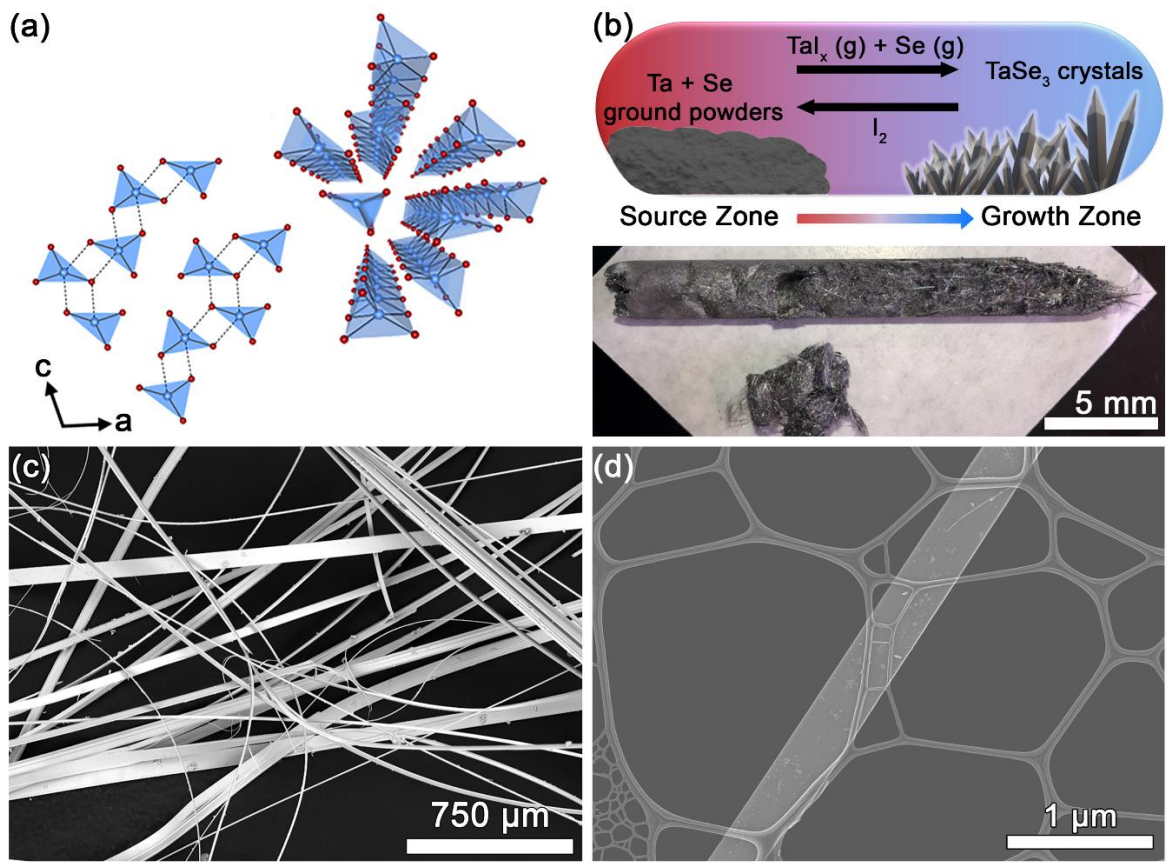

**Figure 1**

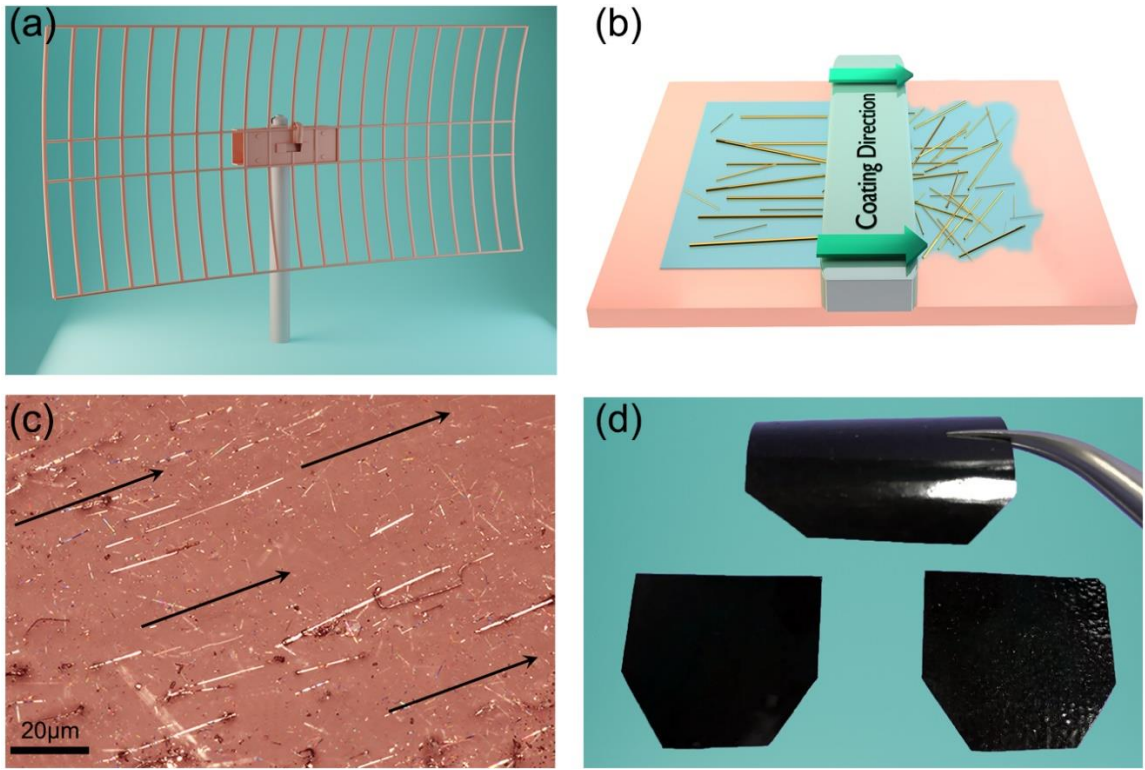

**Figure 2**

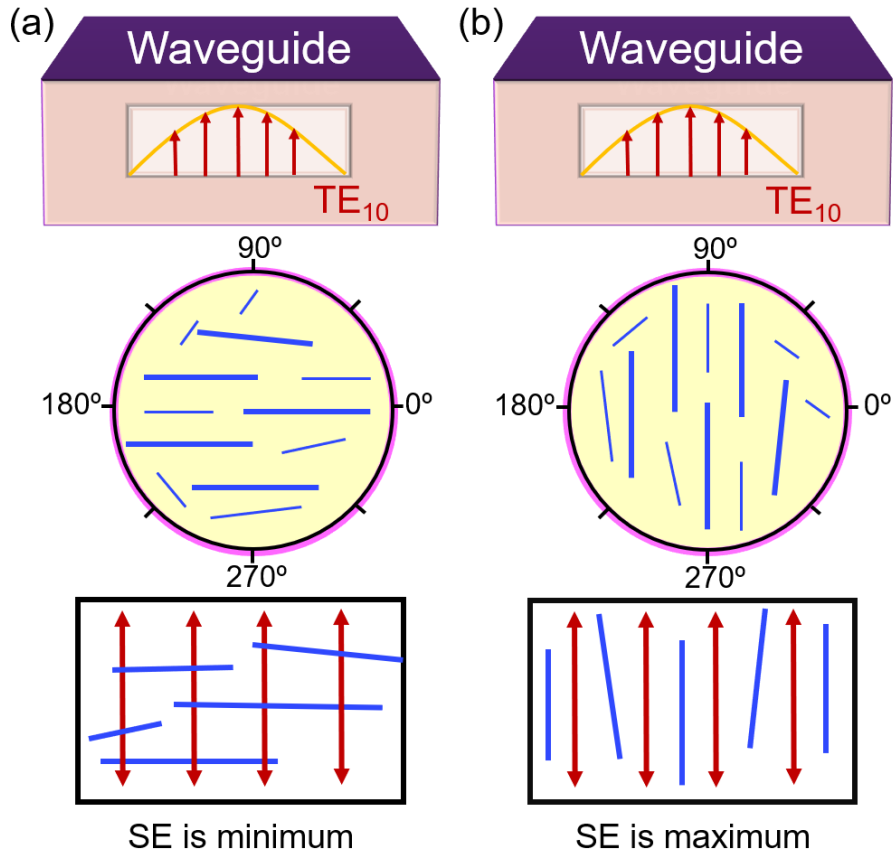

**Figure 3**

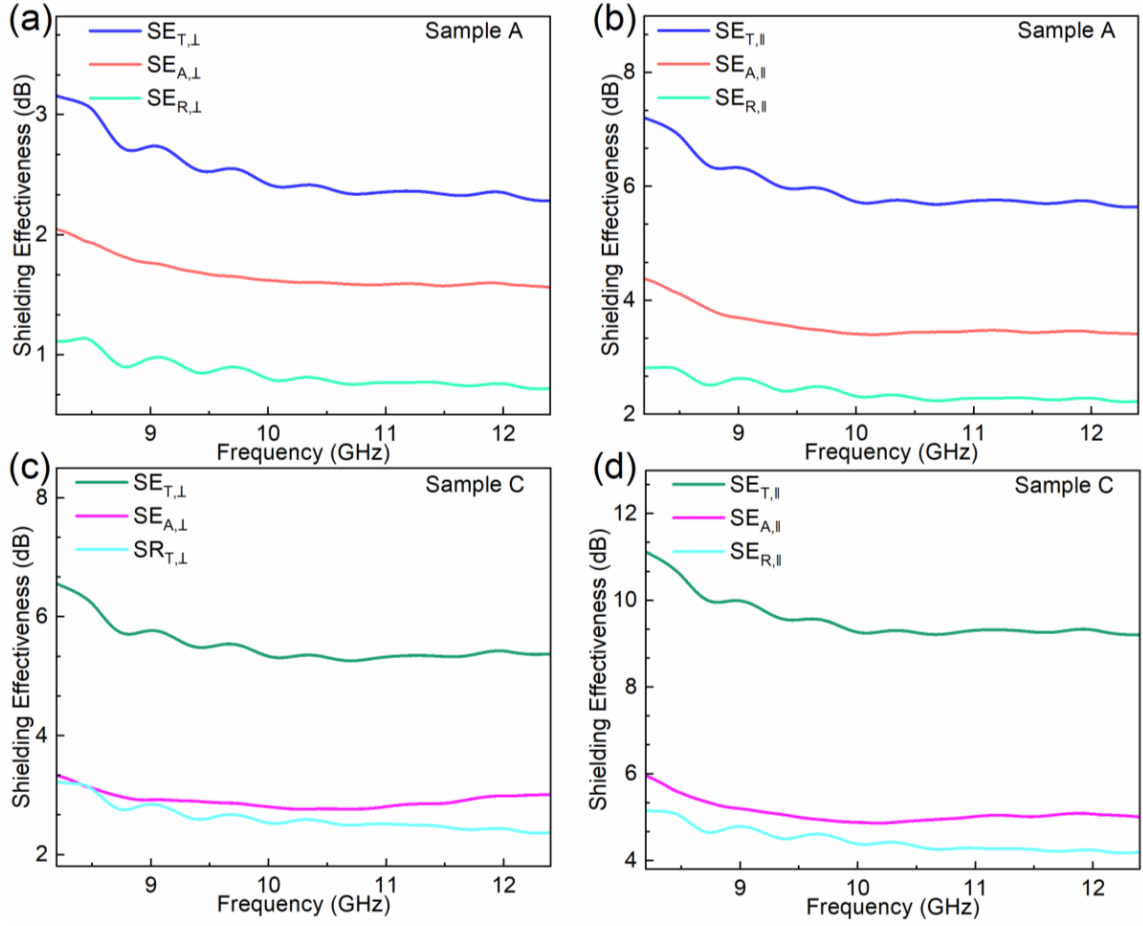

**Figure 4**

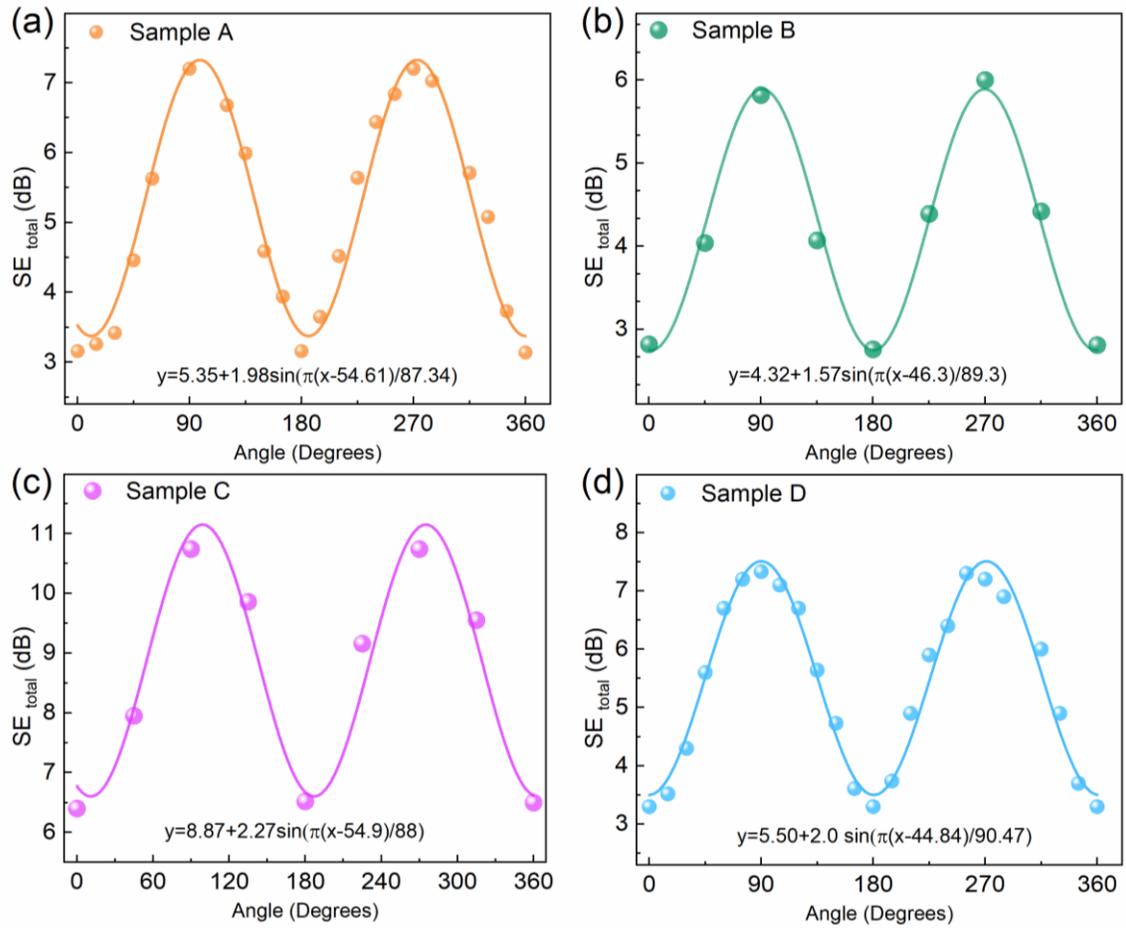

**Figure 5**

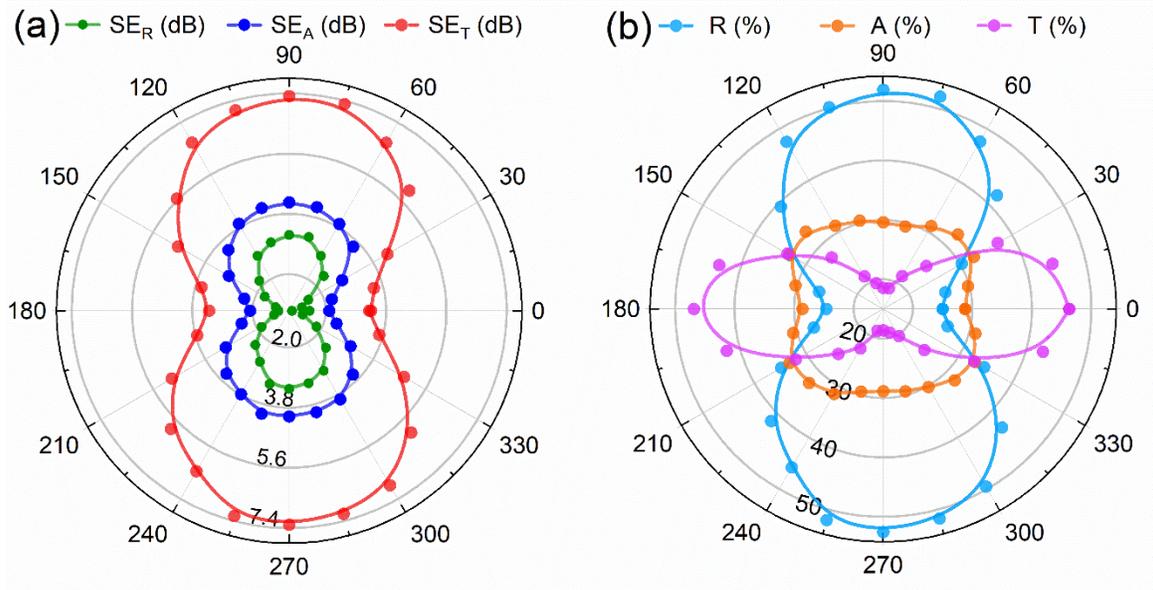

**Figure 6**